\newcommand{\be}{\begin{equation}}
\newcommand{\ee}{\end{equation}}

\def\kmsMpc{\rm{\, km\, s^{-1}\, Mpc^{-1}}}

\documentclass{iau-JDSS-hacked}
\usepackage{graphicx,natbib}

\usepackage{natbib}
\def\reff@jnl#1{{\rm#1\/}}
\def\aj{\reff@jnl{AJ}}                  
\def\araa{\reff@jnl{ARA\&A}}            
\def\apj{\reff@jnl{ApJ}}                
\def\apjl{\reff@jnl{ApJ}}               
\def\apjs{\reff@jnl{ApJS}}              
\def\apss{\reff@jnl{Ap\&SS}}            
\def\aap{\reff@jnl{A\&A}}               
\def\aapr{\reff@jnl{A\&A~Rev.}}         
\def\aaps{\reff@jnl{A\&AS}}             
\def\jcap{\reff@jnl{JCAP}}              
\def\mnras{\reff@jnl{MNRAS}}            
\def\prd{\reff@jnl{Phys.Rev.D}}         
\def\prl{\reff@jnl{Phys.Rev.Lett}}      
\def\pasp{\reff@jnl{PASP}}              
\def\pasj{\reff@jnl{PASJ}}              
\def\nat{\reff@jnl{Nature}}             

\title[] 
{The Hubble constant and new discoveries in cosmology}

\author[] 
{S.~H.~Suyu$^{1,2}$, T.~Treu$^1$, R.~D.~Blandford$^2$,
  W.~L.~Freedman$^3$, S.~Hilbert$^{2}$, C.~Blake$^4$, J.~Braatz$^5$, F.~Courbin$^6$, J.~Dunkley$^7$,
  L.~Greenhill$^8$, E.~Humphreys$^9$, S.~Jha$^{10}$,
  R.~Kirshner$^8$, K.~Y.~Lo$^5$, L.~Macri$^{11}$, B.~F.~Madore$^3$, P.~J.~Marshall$^7$, G.~Meylan$^6$, J.~Mould$^4$, B.~Reid$^{12}$, M.~Reid$^8$, A.~Riess$^{13,14}$,
  D.~Schlegel$^{12}$, V.~Scowcroft$^3$, L.~Verde$^{15}$}

\affiliation{$^1$University of California Santa Barbara, email:
  suyu@physics.ucsb.edu, tt@physics.ucsb.edu, $^2$Kavli Institute for
  Particle Astrophysics and Cosmology, $^3$Carnegie Observatories,
  $^4$Swinburne University of Technology, $^5$National Radio
  Astronomy Observatory, $^6${\'E}cole Polytechnique F{\'e}d{\'e}rale
  de Lausanne, $^7$University of Oxford, $^8$Harvard-Smithsonian
  Center for Astrophysics, $^9$European Southern Observatory,
  $^{10}$Rutgers University, $^{11}$Texas A\&M University,
  $^{12}$Lawrence Berkeley National Laboratory, $^{13}$Johns Hopkins
  University, $^{14}$Space Telescope Science Institute,
  $^{15}$University of Barcelona}

\pubyear{2012}
\volume{Volume 1}  
\pagerange{1-4}
\date{February 20 2012}
\jname{The Hubble Constant: Current and Future Challenges}
\editors{Sherry H.~Suyu, Tommaso Treu, Roger D.~Blandford, Wendy L.~Freedman, ed.}
\begin{document}

\maketitle

\begin{abstract}

We report the outcome of a 3-day workshop on the Hubble constant
($H_0$) that took place during February 6-8 2012 at the Kavli
Institute for Particle Astrophysics and Cosmology, on the campus of
Stanford University\footnote{More details of the 
workshop are given at the website:\\
\indent\indent\indent \texttt{http://web.physics.ucsb.edu/\~{}suyu/H0KIPAC/Home.html}}. The participants met to address the following questions.  Are there
compelling scientific reasons to obtain more precise and more accurate
measurements of $H_0$ than currently available?  If there are, how can
we achieve this goal?  
The answers that emerged from the workshop are (1) better
measurements of $H_0$ provide critical independent constraints on dark energy,
spatial curvature of the Universe, neutrino physics, and validity of
general relativity, (2) a
measurement of $H_0$ to 1\% in both precision and accuracy, supported
by rigorous error budgets, is within reach for several methods, and
(3) multiple paths to independent determinations of $H_0$ are needed
in order to access and control systematics.


\end{abstract}

\begin{figure}[!ht]
\begin{center}
\begin{minipage}[!ht]{0.36\linewidth}
\includegraphics[width=2.5in, clip=true]{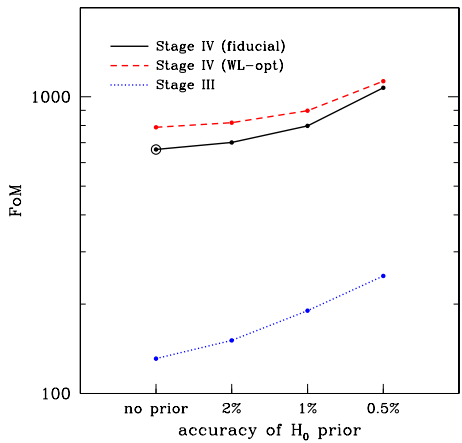}
\end{minipage}\hfill
\begin{minipage}[!ht]{0.50\linewidth}
  \caption{\label{fig:h} Dependence of the FoM from the DETF on the
    accuracy of independent measurements of the Hubble constant that
    would be used as priors in Stage III and IV forecasts from
    \citet{WeinbergEtal12}. The fiducial Stage IV program with
    FoM$=664$ is marked by an open circle.  In all cases, adding a
    prior from an independent measurement of $H_0$ with $\sim$$1\%$
    accuracy increases the FoM by $\sim$$40\%$.  The figure was
    extracted from \citet{WeinbergEtal12}.}
\end{minipage}
\end{center}
\end{figure}

\noindent \textbf{Are there compelling scientific reasons to obtain
  more precise and more accurate measurements of $H_0$ than currently
  available? }
A measurement of the local value of $H_0$ to 1\% precision (i.e., random
errors) {\it and} accuracy (i.e., systematic errors) would provide key
new insights into fundamental physics questions and lead to
potentially revolutionary discoveries. These include the nature of
dark energy and its evolution, the curvature of the Universe as a test
of inflationary models, the mass of neutrinos and the total number of
families of relativistic particles \citep[e.g.,][]{FreedmanMadore10,
  RiessEtal11, WeinbergEtal12, ReidEtal10, SekiguchiEtal10}.  For
example, Figure \ref{fig:h} 
\citep[extracted from][]{WeinbergEtal12} illustrates the dependence of
the Figure of Merit (FoM, for dark energy parameterized by $w(a) =
w_{p} + w_{a} (a_{\rm p}-a)$), introduced by the Dark Energy Task Force
\citep[DETF;][]{AlbrechtEtal06}, on the accuracy of additional
measurements of the Hubble constant that would be used as a prior in Stage
III and Stage IV experiments.  In all cases, adding a prior from an
independent measurement of $H_0$ with accuracy that matches the
uncertainty one would have in the absence of the prior (1.3\% for
Stage III, and 0.7\% for Stage IV) increases the FoM by $\sim$$40\%$.
Measuring $H_0$ therefore raises the value of current Stage III and
future Stage IV experiments for relatively small incremental cost.
In an unparameterized model of $w$ (e.g., in redshift bins), Stage IV
Cosmic Microwave Background (CMB), Baryon Acoustic Oscillation (BAO),
Supernovae (SN) and Weak Lensing (WL) surveys will only constrain
$H_0$ to $\gtrsim 10\%$ \citep{WeinbergEtal12}; thus an independent
measurement of $H_0$ 
provides the only knowledge of the Hubble constant.  This highlights
that probes at $z>0$ (e.g., CMB and BAO) do not measure $H_0$: they
provide \textit{derived} $H_0$ constraints only within a given and
well-specified cosmological model.  In addition, accurate determinations
of distances to objects that are needed for local $H_0$ measurements
provide an invaluable opportunity to learn 
about gravitational physics of an inhomogeneous Universe.  Therefore, a
1\% measurement of $H_0$ in combination with higher redshift probes
provides a fundamental test of the foundations of cosmological models,
including possible departures from general relativity.

\noindent \textbf{How can we achieve 1\% precision and accuracy?}  The
value of $H_0$ has long been controversial because of many systematic
effects in any given method that need to be understood, controlled and
quantified.  Nonetheless, similar to the much better
publicized advances in measuring $\Omega_{\rm m}$ and
$\Omega_{\Lambda}$, ``precision cosmology'' has taken place in
measuring $H_0$ as observations improve, errors are controlled and
independent methods are employed.  We critically assessed the
following range of methods at the workshop, 
focusing on their strengths, weaknesses and prospects for measuring
$H_0$ to $1\%$: Cepheids, Tip of the Red Giant Branch (TRGB), Type Ia
Supernovae, Surface 
Brightness Fluctuations (SBF), Masers, and
Gravitational Lens Time Delays.  Recent reviews on these topics can be
found in \citet{FreedmanMadore10}, \citet{Lo05}, \citet{Treu10}, and
\citet{WeinbergEtal12}.  
All of these methods currently measure $H_0$ to
within $10\%$, finding it to be in the range $68$ to $79\kmsMpc$, a remarkable
consistency given the well-known history of this constant
\citep[e.g.,][]{RiessEtal11, FreedmanEtal11, HislopEtal11, 
  MagerEtal08, BlakesleeEtal10, BraatzEtal10, SuyuEtal10}.  Advances
in reconstructing peculiar velocities \citep[e.g.,][]{LavauxEtal10} should
help reduce the uncertainties in future local measurements of $H_0$.

We summarize the current status of each method below.
\begin{itemize}

\item Recent progress in observing Cepheids in the infrared leads to
  0.1 mag dispersion (5\% in distance) for a single Cepheid.  A 1\%
  $H_0$ measurement should be achievable in principle from this
  technique, and requires a reduction of uncertainties due to crowding
  and metallicity of stars together with an increase in the number of
  calibrators.

\item TRGB can measure 0.1 mag (5\% in distance), which appears to be
  the limit in accuracy.  This approach is restricted to nearby
  galaxies and provides a valuable independent check of Cepheids.

\item Infrared observations of SN Ia yield distances with 10\% precision.  
  These in combination with Cepheid and maser distances lead to a
  $\sim$$3\%$ measurement in $H_0$ \citep{RiessEtal11}.
  The biggest challenge for improvement is the slow increase in the
  number of calibrators nearby enough to detect Cepheids in the host
  (one every 2$-$3 years).  Future work to reduce the uncertainty to
  1\% includes characterizing the population, studying the dependence
  in metallicity and understanding the theory of SN Ia explosions.

\item Recent SBF study by \citet{BlakesleeEtal10} suggests a scatter
  of only 1.5\% in distance for a   
  small sample; the approach is promising but requires more study to
  quantify biases and systematics.

\item There are currently $\sim$$10$ known maser galaxies whose
  geometric distances can be measured at the $\sim$$10\%$ level.  An
  updated distance measurement to NGC 4258 with 2.5\% precision was
  presented.  Getting to the 1\% level in $H_0$ would need both
  discovery of more systems (NGC 4258 is unfortunately a rare nearby
  system), which would be challenging before the SKA, and better
  modeling/understanding of disk physics.  A factor of $\sim$$2$
  reduction in uncertainty for NGC 4258 and an identification of a
  second NGC-4258-like maser in a galaxy with detectable Cepheids are
  both valuable as they would advance calibrations of other indicators.

\item Gravitational lens time delays can measure time-delay distances
  to $\sim$$5\%$ for a single lens system.  The distance is primarily
  sensitive to $H_0$, and the $5\%$ translates to $\sim$$7\%$ in $H_0$
  assuming uniform prior distributions of $\Omega_{\Lambda}$, $w$ and $H_0$.
  A comparison of multiple systems is crucial to test for any residual
  systematic errors in the correction of the line of sight effects.
  There are currently tens of time-delay lenses from
  Cosmograil\footnote{COSmological MOnitoring of GRAvItational
    Lenses\ \ \ \ \ \ \ \ \ \ \ \ \ \ \ \ \ \ \ \ \ \ \ \ \ \ \ \ \ \
    \ \ \ \ \ \ \ \ \ \ \ \ \ \ \ \ \ \ \ \ \ \ }, and future samples of hundreds make 1\% an achievable
  goal when systematics are under control.

\end{itemize}

We also investigated CMB and BAO, both of which do not measure
directly $H_0$, but can predict it within the context of a model.
Upcoming CMB data (e.g., from Planck) should be able to break the
degeneracy between the number of relativistic neutrino species and
$H_0$ that is present in current CMB data.  However, an independent
measurement of $H_0$ is needed to extract from the CMB information on
dark energy and spatial curvature of the Universe.  BAO provides a
standard ruler on the sky, but will be limited by cosmic variance at
low $z$.  Measurements of $D_{\rm A}$ and $H(z)$ at $\sim$$1\%$ in various
bins of $z$ appear feasible with current and planned large-scale BAO
surveys (e.g., BOSS and BigBOSS).  Independent analyses using
different approaches by separate groups and independent galaxy
samples probing various galaxy populations are needed to offer
consistency checks and tests for any unknown systematic uncertainties.
We also examined differential ages of cosmic chronometers (old passive
elliptical galaxies) as a probe of $H(z)$.  At present, this technique
produces measurements of $H$ at $z\sim 0$ with a 6\% uncertainty and
of $H(z)$ up to $z\lesssim 1$ with a $\sim$$13\%$ uncertainty. It
remains to be explored if the method can achieve 1\% in accuracy with
targeted data.

\ 

\noindent \textbf{Concluding Remarks}

All approaches can gain by both developing novel analysis techniques
and increasing the sample size (though the $\sqrt{N}$ gain is only
obtained when systematics are under control). A good physical
understanding is highly desirable for all methods (progenitors
of supernovae, structure/dynamics of water masers, etc.).  As most of
the methods are limited by systematics, it becomes ever more important
to provide rigorous error budgets for each approach, make data
available to others, and perform cross checks via, e.g., public
challenges and blind tests that would substantiate the independence.
Furthermore, several distance indicators with comparable 1\% precision
are required to provide a robust estimate of $H_0$ to 1\%
accuracy. Each of the above-mentioned methods is currently worth
pursuing, and extensive work is needed in the next several years to
beat down the systematics to these levels; it is not clear that all
methods will achieve it.  The multiple paths of independent $H_0$
determination are crucial for understanding whether the current
indications of tension between direct local measurements of $H_0$ and
BAO/CMB are signaling new physics.

\bibliographystyle{apj}
\bibliography{H0Communique.bib}

\begin{thebibliography}{15}
\expandafter\ifx\csname natexlab\endcsname\relax\def\natexlab#1{#1}\fi

\bibitem[{{Albrecht} {et~al.}(2006){Albrecht}, {Bernstein}, {Cahn}, {Freedman},
  {Hewitt}, {Hu}, {Huth}, {Kamionkowski}, {Kolb}, {Knox}, {Mather}, {Staggs},
  \& {Suntzeff}}]{AlbrechtEtal06}
{Albrecht}, A., {Bernstein}, G., {Cahn}, R., {Freedman}, W.~L., {Hewitt}, J.,
  {Hu}, W., {Huth}, J., {Kamionkowski}, M., {Kolb}, E.~W., {Knox}, L.,
  {Mather}, J.~C., {Staggs}, S., \& {Suntzeff}, N.~B. 2006, ArXiv e-prints
  (astro-ph/0609591)

\bibitem[{{Blakeslee} {et~al.}(2010){Blakeslee}, {Cantiello}, {Mei},
  {C{\^o}t{\'e}}, {Barber DeGraaff}, {Ferrarese}, {Jord{\'a}n}, {Peng},
  {Tonry}, \& {Worthey}}]{BlakesleeEtal10}
{Blakeslee}, J.~P., {Cantiello}, M., {Mei}, S., {C{\^o}t{\'e}}, P., {Barber
  DeGraaff}, R., {Ferrarese}, L., {Jord{\'a}n}, A., {Peng}, E.~W., {Tonry},
  J.~L., \& {Worthey}, G. 2010, \apj, 724, 657

\bibitem[{{Braatz} {et~al.}(2010){Braatz}, {Reid}, {Humphreys}, {Henkel},
  {Condon}, \& {Lo}}]{BraatzEtal10}
{Braatz}, J.~A., {Reid}, M.~J., {Humphreys}, E.~M.~L., {Henkel}, C., {Condon},
  J.~J., \& {Lo}, K.~Y. 2010, \apj, 718, 657

\bibitem[{{Freedman} \& {Madore}(2010)}]{FreedmanMadore10}
{Freedman}, W.~L. \& {Madore}, B.~F. 2010, \araa, 48, 673

\bibitem[{{Freedman} {et~al.}(2011){Freedman}, {Madore}, {Scowcroft}, {Monson},
  {Persson}, {Seibert}, {Rigby}, {Sturch}, \& {Stetson}}]{FreedmanEtal11}
{Freedman}, W.~L., {Madore}, B.~F., {Scowcroft}, V., {Monson}, A., {Persson},
  S.~E., {Seibert}, M., {Rigby}, J.~R., {Sturch}, L., \& {Stetson}, P. 2011,
  \aj, 142, 192

\bibitem[{{Hislop} {et~al.}(2011){Hislop}, {Mould}, {Schmidt}, {Bessell}, {Da
  Costa}, {Francis}, {Keller}, {Tisserand}, {Rapoport}, \&
  {Casey}}]{HislopEtal11}
{Hislop}, L., {Mould}, J., {Schmidt}, B., {Bessell}, M.~S., {Da Costa}, G.,
  {Francis}, P., {Keller}, S., {Tisserand}, P., {Rapoport}, S., \& {Casey}, A.
  2011, \apj, 733, 75

\bibitem[{{Lavaux} {et~al.}(2010){Lavaux}, {Tully}, {Mohayaee}, \&
  {Colombi}}]{LavauxEtal10}
{Lavaux}, G., {Tully}, R.~B., {Mohayaee}, R., \& {Colombi}, S. 2010, \apj, 709,
  483

\bibitem[{{Lo}(2005)}]{Lo05}
{Lo}, K.~Y. 2005, \araa, 43, 625

\bibitem[{{Mager} {et~al.}(2008){Mager}, {Madore}, \& {Freedman}}]{MagerEtal08}
{Mager}, V.~A., {Madore}, B.~F., \& {Freedman}, W.~L. 2008, \apj, 689, 721

\bibitem[{{Reid} {et~al.}(2010){Reid}, {Verde}, {Jimenez}, \&
  {Mena}}]{ReidEtal10}
{Reid}, B.~A., {Verde}, L., {Jimenez}, R., \& {Mena}, O. 2010, \jcap, 1, 3

\bibitem[{{Riess} {et~al.}(2011){Riess}, {Macri}, {Casertano}, {Lampeitl},
  {Ferguson}, {Filippenko}, {Jha}, {Li}, \& {Chornock}}]{RiessEtal11}
{Riess}, A.~G., {Macri}, L., {Casertano}, S., {Lampeitl}, H., {Ferguson},
  H.~C., {Filippenko}, A.~V., {Jha}, S.~W., {Li}, W., \& {Chornock}, R. 2011,
  \apj, 730, 119

\bibitem[{{Sekiguchi} {et~al.}(2010){Sekiguchi}, {Ichikawa}, {Takahashi}, \&
  {Greenhill}}]{SekiguchiEtal10}
{Sekiguchi}, T., {Ichikawa}, K., {Takahashi}, T., \& {Greenhill}, L. 2010,
  \jcap, 3, 15

\bibitem[{{Suyu} {et~al.}(2010){Suyu}, {Marshall}, {Auger}, {Hilbert},
  {Blandford}, {Koopmans}, {Fassnacht}, \& {Treu}}]{SuyuEtal10}
{Suyu}, S.~H., {Marshall}, P.~J., {Auger}, M.~W., {Hilbert}, S., {Blandford},
  R.~D., {Koopmans}, L.~V.~E., {Fassnacht}, C.~D., \& {Treu}, T. 2010, \apj,
  711, 201

\bibitem[{{Treu}(2010)}]{Treu10}
{Treu}, T. 2010, \araa, 48, 87

\bibitem[{{Weinberg} {et~al.}(2012){Weinberg}, {Mortonson}, {Eisenstein},
  {Hirata}, {Riess}, \& {Rozo}}]{WeinbergEtal12}
{Weinberg}, D.~H., {Mortonson}, M.~J., {Eisenstein}, D.~J., {Hirata}, C.,
  {Riess}, A.~G., \& {Rozo}, E. 2012, ArXiv e-prints (1201.2434)

\end{thebibliography}

\end{document}